\documentclass[twocolumn]{aastex631}

\newcommand{\logg}{\ensuremath{\log g}}
\newcommand{\Teff}{\ensuremath{T_{\textrm{eff}}}}
\newcommand{\teff}{\Teff}
\newcommand{\abunratio}[2]{\ensuremath{{[\mathrm{#1}/\mathrm{#2}]}}}
\newcommand{\feh}{\abunratio{Fe}{H}}

\newcommand{\afe}{\abunratio{\alpha}{Fe}}
\newcommand{\kms}{\ensuremath{\textrm{km}~\textrm{s}^{-1}}}

\newcommand{\vmicro}{$\xi$}

\newcommand{\masyr}{\ensuremath{\textrm{mas}~\textrm{yr}^{-1}}}
\newcommand{\dex}{\ensuremath{\textrm{dex}}}

\newcommand{\acronym}[1]{{\small{#1}}}
\newcommand{\Msun}{\ensuremath{\textrm{M}_{\odot}}}

\newcommand{\project}[1]{\textsl{#1}}
\newcommand{\gaia}{\project{Gaia}}

\newcommand{\thepayne}{\project{The~Payne}}

\newcommand{\numstar}{10}

\usepackage{graphicx}
\usepackage{amssymb}
\usepackage{amsmath}
\usepackage{times}

\usepackage{subfigure}
\usepackage[T1]{fontenc}
\UseRawInputEncoding

\begin{document}


\title[Chemistry of the Orphan Stream in APOGEE DR17]{On the Hunt for the Origins of the Orphan--Chenab Stream: Detailed Element Abundances with APOGEE and Gaia}

\newcommand{\affut}{Department of Astronomy, The University of Texas at Austin, 2515 Speedway Boulevard, Austin, TX 78712, USA}
\newcommand{\affcca}{Center for Computational Astrophysics, Flatiron Institute,162 Fifth Avenue, New York, NY 10010, USA}

\newcommand{\affRlane}{Centro de Investigaci\'on en Astronom\'ia, Universidad Bernardo O'Higgins, Avenida Viel 1497, Santiago, Chile}

\newcommand{\affAntof}{Centro de Astronom{\'i}a (CITEVA), Universidad de Antofagasta, Avenida Angamos 601, Antofagasta 1270300, Chile}

\author[0000-0002-1423-2174]{Keith~Hawkins}
\affiliation{\affut}
\email{keithhawkins@uteaxs.edu}
\correspondingauthor{Keith Hawkins}

\author[0000-0003-0872-7098]{Adrian~M.~Price-Whelan}
\affiliation{\affcca}

\author[0000-0003-2178-8792]{Allyson~A.~Sheffield}
\affiliation{CUNY - LaGuardia Community College, Department of Natural Sciences, Long Island City, NY 11101, USA}

\author[0000-0002-0837-6331]{Aidan~Z.~Subrahimovic}
\affiliation{Center for Cosmology and Particle Physics, New York University, Department of Physics, 726 Broadway, New York, NY 10003, USA}

\author{Rachael~L.~Beaton}
\affiliation{The Observatories of the Carnegie Institution for Science, 813 Santa Barbara Street, Pasadena, CA 91101, USA}

\author{Vasily~Belokurov}
\affiliation{\affcca}
\affiliation{Institute of Astronomy, Madingley Rd, Cambridge, CB3 0HA, UK}


\author[0000-0002-8448-5505]{Denis Erkal}
\affiliation{Department of Physics, University of Surrey, Guildford GU2 7XH, UK}

\author[0000-0003-2644-135X]{Sergey E. Koposov}
\affiliation{Institute for Astronomy, University of Edinburgh, Royal Observatory, Blackford Hill, Edinburgh EH9 3HJ, UK}
\affiliation{Institute of Astronomy, University of Cambridge, Madingley Road, Cambridge CB3 0HA, UK}

\author[0000-0003-1805-0316]{Richard~R.~Lane}
\affiliation{\affRlane}

\author{Chervin~F.~P.~Laporte}
\affiliation{Institut de Ciencies del Cosmos (ICCUB), Universitat de Barcelona (IEEC-UB), Mart\'ı i Franqu\'es 1, E08028 Barcelona, Spain}

\author[0000-0003-4752-4365]{Christian~Nitschelm}
\affiliation{\affAntof}




\label{firstpage}

\begin{abstract}
Stellar streams in the Galactic halo are useful probes of the assembly of galaxies like the Milky Way. Many tidal stellar streams that have been found in recent years are accompanied by a known progenitor globular cluster or dwarf galaxy. However, the Orphan--Chenab (OC) stream is one case where a relatively narrow stream of stars has been found without a known progenitor. In an effort to find the parent of the OC stream, we use astrometry from the early third data release of ESA's \gaia\ mission (\gaia~EDR3) and radial velocity information from the SDSS-IV APOGEE survey to find up to 13 stars that are likely members of the OC stream. We use the APOGEE survey to study the chemical nature (for up to \numstar~stars) of the OC stream in the $\alpha$ (O, Mg, Ca, Si, Ti, S), odd-Z (Al, K, V), Fe-peak (Fe, Ni, Mn, Co, Cr) and neutron capture (Ce) elemental groups. We find that the stars that make up the OC stream are not consistent with a mono-metallic population and have a median metallicity of --1.92~dex with a dispersion of 0.28~dex. Our results also indicate that the $\alpha$-elements are depleted compared to the known Milky Way populations and that its [Mg/Al] abundance ratio is not consistent with second generation stars from globular clusters. The detailed chemical pattern of these stars indicates that the OC stream progenitor is very likely to be a dwarf spheroidal galaxy with a mass of $\sim10^6$~\Msun. 

\end{abstract}


\section{Introduction}
\label{sec:Introduction}

In $\Lambda$-cold dark matter ($\Lambda$CDM) cosmology, it is believed that galaxies in general are assembled in a hierarchical  way through the accretion of small sub-Galactic systems to construct larger ones and eventually systems over a broad range in mass and size \citep[e.g.,][]{Searle1978, Davis1985, Bullock2005}. In this context, one would expect the stellar halo of the Milky Way to be built largely from the accretion of smaller objects, as well as a smaller component of material formed within its viral radius \citep[so called in situ material, e.g., ][and references therein]{Helmi2000, Cooper2015, Hawkins2015, Bonaca2017}.

The Galactic halo contains many relic substructures in the form of gravitationally-bound clusters and dwarf galaxies and disrupted analogs in the form of tidal stellar streams from both globular clusters and dwarf galaxies alike. The modern ``field of streams'' \citep{Belokurov2006, Bonaca:2012} in our stellar halo demonstrates the importance of accretion processes in the build up of stellar halos. Studying the amount and properties of substructure in the Milky Way's stellar halo has provided an important way of constraining the importance of accretion processes in forming stellar halos \citep[e.g.,][]{Naidu:2020}, of measuring the mass and profile of dark matter in the Galaxy \citep[e.g.][]{Majewski2003, Law2010, Koposov2010, Gibbons2014}, and of constraining the small-scale properties of dark matter within galaxies \citep[e.g.,][]{Carlberg:2012,Erkal:2016,Price-Whelan:2018, Bonaca:2019, Banik:2021}. 

In this work, we focus on the Orphan--Chenab (OC) Stream: The Orphan stream was discovered independently by two separate teams \citep{Grillmair2006, Belokurov2007} and the Chenab Stream was discovered photometrically as a roughly 19$^{\circ}$ stellar feature in the Dark Energy Survey DR1 \citep{shipp18}, in the same direction on the sky as the Southern Galactic component of the Orphan Stream and at an estimated photometric distance of 40 kpc. The spatial overlap between the Orphan and Chenab streams was shown in the all-sky view presented by \cite{Koposov2019}. Additionally, both \cite{Koposov2019} and \cite{Shipp19} illustrated that the proper motion signals of Orphan and Chenab streams are consistent. The equivalence of the Orphan and Chenab streams across multiple kinematic parameters is now understood as strong evidence that these streams are both remnants of the tidal disruption of the same progenitor. In particular, a misalignment in the northern/southern stream poles can be explained by a dynamic encounter of the Southern portion of the OC stream with the Large Magellanic Cloud \citep{Erkal2019}. 

Despite being a large structure traced over $\sim 210^\circ$ on the sky (or $\sim 150~\textrm{kpc}$ in physical length; \citealt{Grillmair2015, Fardal2019, Koposov2019}), the OC stream still has no known progenitor system. The OC stream is also relatively narrow, just $\sim 2^{\circ}$ wide on the sky, but its width is significantly broader than known streams associated with globular cluster systems (e.g., the Palomar 5 stream, \citealt{Odenkirchen2003,Erkal2017,Bonaca:2020}). However, the stellar mass of the OC stream implies that the system that it originated in could be on the massive end of classical globular clusters \cite[e.g.][]{Grillmair2006,Belokurov2007}. The stream is also relatively low in surface brightness and has a distance range from $\sim$19--55~kpc (there is a steep distance gradient along the stream; \citealt{Sesar2013, Koposov2019}. Recent orbital fitting of the stream indicates that it has a Galactocentric pericenter of $\sim$16~kpc and an apocenter of $\sim$90~kpc\citep[e.g.][]{Newberg2010}, which implies a very eccentric orbit.

There are several theories for potential progenitors to the OC stream. The first is that the OC stream originates from a globular cluster system. \cite{Koposov2019} has found that as many as  7 globular clusters are within 7$^{\circ}$ of the stream's great circle (see the top panel of their Fig.~14). Of these globular clusters, two have been discussed in the literature as potential parents for the Orphan stream, namely NGC~2419 \citep[discussed in][]{Bruns2011} and Ruprecht~106 \citep[discussed in][]{Grillmair2015, Koposov2019}. However, both are thought implausible based on chemo-kinematic arguments \citep[e.g.][]{Casey2014, Grillmair2015}. Recently, \cite{Li2022} noted that the globular cluster system Laevens~3 \citep{Laevens2015}, may be a potential progenitor of OC stream. The chemical exploration of the OC stream in this study will enable us to further test this idea.  Another possibility is that the OC stream has originated from an ultra faint dwarf galaxy, such as Segue~1 \citep[e.g.,][]{Gilmore2013, Casey2014} or Ursa~Major~II \citep[e.g.,][]{Fellhauer2007} or Grus~II \citep[e.g.,][]{Koposov2019}. Distinguishing between these theories of the long lost parent of the OC stream is, in some sense, the great challenge. The kinematic, spatial {\it and} chemical nature of the stream is important to quantify in order to better understand where it came from \citep[e.g.,][]{Li2022, Naidu2022}. Specifically, the chemical fingerprint of the OC stream, which has been relatively less explored than its kinematic signature, will offer a powerful tool in our hunt for the parent system \citep[as it has for other systems, e.g.,][]{Aguado2021,Carrillo2022, Matsuno2022}.

Therefore, in this work, we revisit the chemical properties of the Orphan--Chenab stream with the newest data release (DR17) of the Sloan Digital Sky Survey's \citep{Gunn2006} Apache Point Observatory Galactic Evolution Experiment DR17 data \citep[APOGEE,][]{Nidever2015, SDSSDR17,Majewski2017,Blanton:2017, Wilson2019} survey in order to ascertain a possible parent population and provide chemical abundances for the largest sample of OC stars to-date.  To do this, we begin in section~\ref{sec:data}, with a description of the exquisite data from the early third data release (EDR3) from the \gaia\ mission \citep{Gaia:edr3}, which we use in order to select probable members of the OC stream (section~\ref{subsec:selection}). We also describe the spectroscopic data from the APOGEE survey, where the stellar parameter and chemical abundance information for the OC stars are sourced from (section~\ref{subsec:APOGEE}). In section~\ref{sec:results}, we describe the chemical abundance patterns for stars in the OC stream and discuss these results in the context of probable parent populations in section~\ref{sec:discussion}. We end by summarizing in section~\ref{sec:summary}.

\section{Data} \label{sec:data}

We use data from the early Data Release 3 (EDR3) of the European Space Agency's
\gaia\ mission \citep{Gaia:edr3,Gaia2016} cross-matched to Data Release 17
(DR17) of the APOGEE survey \citep[][using the cross-match provided by the
APOGEE team and included in the DR17 data
files]{SDSSDR17,Majewski2017,Blanton:2017}.
We use the astrometric data from \gaia\ to select members of the OC
stream based on their kinematics, and use the element abundance measurements
from the APOGEE survey to characterize the chemical properties of the stream.


\subsection{Selection of Probable Orphan Stream Stars}
\label{subsec:selection}

Though candidate members of the OC stream were explicitly targeted
in certain fields in the APOGEE-2 survey \citep[ORPHAN-1 through
ORPHAN-5][]{Zasowski:2017}, we search all APOGEE fields to identify candidate
members with the hope of finding serendipitously-targeted stream members.
We start by selecting all APOGEE sources that match a set of expected stellar
parameters for red giant branch (RGB) stars in the stellar halo.
Based on the known distance trend of the stream from precise RR Lyrae (RRL)
distance measurements \citep{Koposov2019}, we expect that only mid- to upper-RGB
stars will be luminous enough to have been observed in the $H$-band by APOGEE,
so we first select only stars with surface gravities between $-0.5 < \log g < 2$
(with the lower limit primarily to remove missing values of $-9999$).
We then transform the sky positions of the remaining APOGEE sources from
equatorial coordinates to the OC stream coordinate frame defined in
\cite{Koposov2019} and implemented in the \texttt{Gala} package
\citep{gala:2017} to utilize the \texttt{astropy.coordinates} transformation
framework \citep{astropy:2018}.

\begin{figure*}[ht]
	 \includegraphics[width=\textwidth]{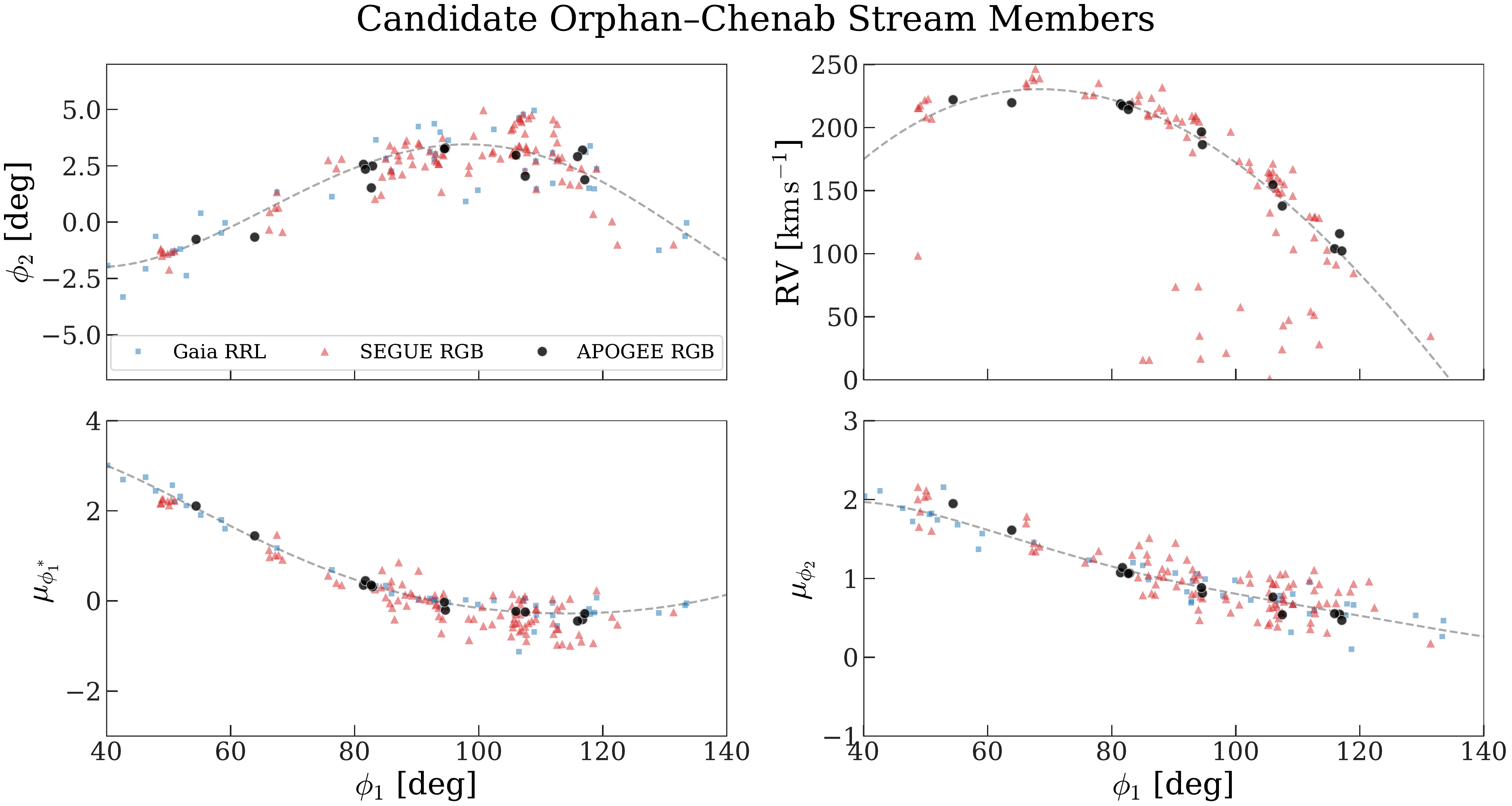}
	\caption{
		In all panels where values are available, blue square markers show the
		OC stream RRL stars \citep{Koposov2019} and red triangles
		show the stream RGB stars \citep{Koposov2019} cross-matched to the
		\acronym{SEGUE} survey.
		Sky positions $\phi_1, \phi_2$ (in OC stream-aligned
		coordinates; \citealt{Koposov2019}) and proper motion components
		$\mu_{\phi_1^*}, \mu_{\phi_2}$ are all taken from cross-matches with
		\gaia\ EDR3.
		Radial velocity (RV) measurements come from either APOGEE or
		\acronym{SEGUE}.
		The dashed lines show 5th- or 3rd-order polynomial fits to the RRL (for
		the $\phi_2$, $\mu_{\phi_1^*}$, and $\mu_{\phi_2}$ panels) or RGB (for
		the RV panel) stars as a function of stream longitude $\phi_1$.
		The black circle markers show the 13 stars identified as stream members
		from the APOGEE DR17 catalog as described in
		Section~\ref{subsec:selection}.
	}
	\label{fig:members}
\end{figure*}

To define criteria for selecting stream stars in sky position, proper motion
components, and radial velocity, we use the candidate RRL and RGB star members
of the OC stream from \cite{Koposov2019}.
In detail, as a function of OC stream longitude $\phi_1$, we fit
5th-order polynomials to the sky position in OC stream latitude
$\phi_2$, the proper motion in stream longitude $\mu_{\phi_1^*}$ (including the
$\cos\phi_2$ term), and the proper motion in stream latitude $\mu_{\phi_2}$
using the OC RRL stars.
Using these polynomial tracks, we select all APOGEE stars within
$\pm2.5^\circ$ of the RRL track in $\phi_2$, within $\pm0.75~\masyr$ of the RRL
track in $\mu_{\phi_1^*}$, and within $\pm0.5~\masyr$ of the RRL track in
$\mu_{\phi_2}$, where these widths were chosen to span the maximum dispersion of
the RRL stars in each dimension.
We then cross-match the \cite{Koposov2019} RGB stream member stars to the
\acronym{SEGUE} survey \cite{Yanny2009} and fit a 3rd-order polynomial to RGB
stars with $v_{\textrm{helio}} > 100~\kms$ (to remove contamination; see upper
right panel of Figure~\ref{fig:members}) as a function of $\phi_1$.
We finally select APOGEE stars with velocities within $\pm 20~\kms$ (about
5-$\sigma$, using the velocity dispersion of the stream as measured in
\citealt{Li2022}) of the RGB track in radial velocity.

Figure~\ref{fig:members} summarizes our kinematic selection of OC
stream members along with our final sample of APOGEE DR17 RGB stars.
The styled markers in the figure show the three different datasets used here:
The RRL and RGB candidate stream members from \cite{Koposov2019} are shown as
blue (square) and red (triangle) markers, and the final sample of 13 APOGEE
stars that are likely members of the OC stream are shown with the
black (circle) markers.

\subsection{Chemical Abundances from the APOGEE Survey} \label{subsec:APOGEE}
The APOGEE spectroscopic survey has collected more than 700,000 moderate resolution ($\textrm{R} = \lambda/\Delta\lambda \sim 22,500$) near infrared (H-band, 1.51--1.60~$\mu\textrm{m}$) spectra of mostly red giant stars across the Milky Way. The primary goal of the APOGEE survey is to study the kinematic and chemical properties of stars across the Milky Way to better understand its structure and nature. In recent years, the survey has also expanded to include red giant stars deep in the stellar halo and Galactic bulge in combination with the many stars it has already observed in the Galactic disk. The survey uses both the 2.5m SDSS telescope at the Apache Point Observatory in the Northern hemisphere and the 2.5m du Pont telescope in the Southern hemisphere. 

The spectra obtained by the APOGEE survey have enabled the exploration of the detailed chemical properties of the Milky Way and its substructures. The APOGEE Stellar Parameter and Abundance Pipeline \citep[ASPCAP,][]{GarciaPerez2015} is the primary tool that has been used to derive both the atmospheric parameters (\teff, \logg, \feh, microturblent velocity, \vmicro) and chemical abundances for up to 20 elements \citep[e.g.][]{Holtzman2015, Holtzman2018,SDSSDR17}. These elements span the $\alpha$ (O, Mg, Ca, Si, Ti, S), odd-Z (Al, K V) Fe-peak (Fe, Ni, Mn, Co, Cr), and neutron capture (Ce) groups. Currently, ASPCAP also allows for derivation of heavier elements including s-process elements Nd, Ce, Rb, and Yb \citep[e.g.][]{Hawkins2016b, Hasselquist2016, Cunha2017}, however only a relatively small fraction contain measurements of these elements from the current DR17 analysis \citep{Holtzman2018, SDSSDR17,Majewski2017,Blanton:2017}. While we provide a short description here, we refer the reader to \cite{GarciaPerez2015} and Jonsson et al. (in prep.), for a thorough description of ASPCAP. In order to measure the stellar properties, ASPCAP (as setup for APOGEE DR17 and for these data presented here) begins by first deriving the atmospheric parameters (\teff, \logg, \feh) of the stars along with the [C/H] and [N/H] \afe\ values. This is done finding a best matching spectrum (via $\chi^2$ minimization) within an interpolated grid of synthetic spectra using the FERRE code \citep{AllendePrieto2006, Zamora2015}. The \vmicro\ parameter was determined through an empirically determine relationship with \logg. 


Once this is complete, the ASPCAP pipeline derives stellar abundances by fitting models to the observed spectra in specific windows. Each window is weighted by its sensitivity to a given abundance and the how well the line modeled in Arcturus can be reproduced. Uncertainties in the stellar abundances are estimated using the scatter of each abundance within (assumed) chemically homogeneous open clusters. We refer the reader to the APOGEE data release papers \citep{Holtzman2015,Holtzman2018} for more details on the exact procedures.


In recent years, the OC stream was specifically targeted within the SDSS-IV APOGEE-2 survey. In order to obtain reliable stellar parameters and chemical abundances, we take the probable OC stream stars observed in APOGEE (described in sec.~\ref{subsec:selection}) and apply several quality control cuts. Following other studies of distant systems \citep[e.g.,][]{Hasselquist2021}, we begin by selecting stars that have spectra with signal-to-noise ratios (SNR) of at least 70. We also remove stars that had the \textsc{STAR\_BAD} bit set in the \textsc{ASPCAPFLAG}. These two criteria were employed to remove problematic or suspect stellar parameters and derived abundances, while attempting to keep the largest number of measurements. As noted in the APOGEE data release papers \citep[e.g.][]{GarciaPerez2015, Holtzman2015, Holtzman2018,SDSSDR17}, this is done in order ensure that the ASPCAP pipeline converged with no major warnings (e.g., if there were known issues with the \teff, \logg, $\chi^2$, rotation, SNR, or if the derived parameters were near a grid edge, etc., the \textsc{STAR\_BAD} flag would be set) indicating that the reported parameters and abundances may not be reliable. This cut reduced the sample from 13 probably OC members to 10. We also require that $3500 < \teff < 5500~\textrm{K}$, $\logg > 0.5~\dex$, but all remaining OC member stars pass this selection. We note here that an additional 4 stars would be removed if we applied the more strict quality control cut that requires \textsc{ASPCAPFLAG} to equal 0 (i.e. absolutely no issues were raised in the \textsc{ASPCAPFLAG} pipeline). Even though the sample would be reduced, the results would remain unchanged. 

In addition to the OC stream stars, for comparison, we also source the chemical abundance information of several globular clusters \citep[specifically, we take members for the M107, M13, M71, M92, and M15 clusters from][]{Meszaros2020} and dwarf galaxies \citep[we take members of the Sagittarius dwarf galaxy from][]{Hasselquist2021}. These comparisons are on the same abundance scale as the OC stream stars and therefore give us a way to determine which chemical patterns are most similar.

\section{Results: The Chemical Composition of Orphan-Chenab Stream Stars} \label{sec:results}

\subsection{$\alpha$ elements (O, Mg, Ca, Si, S, Ti)} \label{subsec:alpha}

\begin{figure*}
	 \includegraphics[width=2\columnwidth]{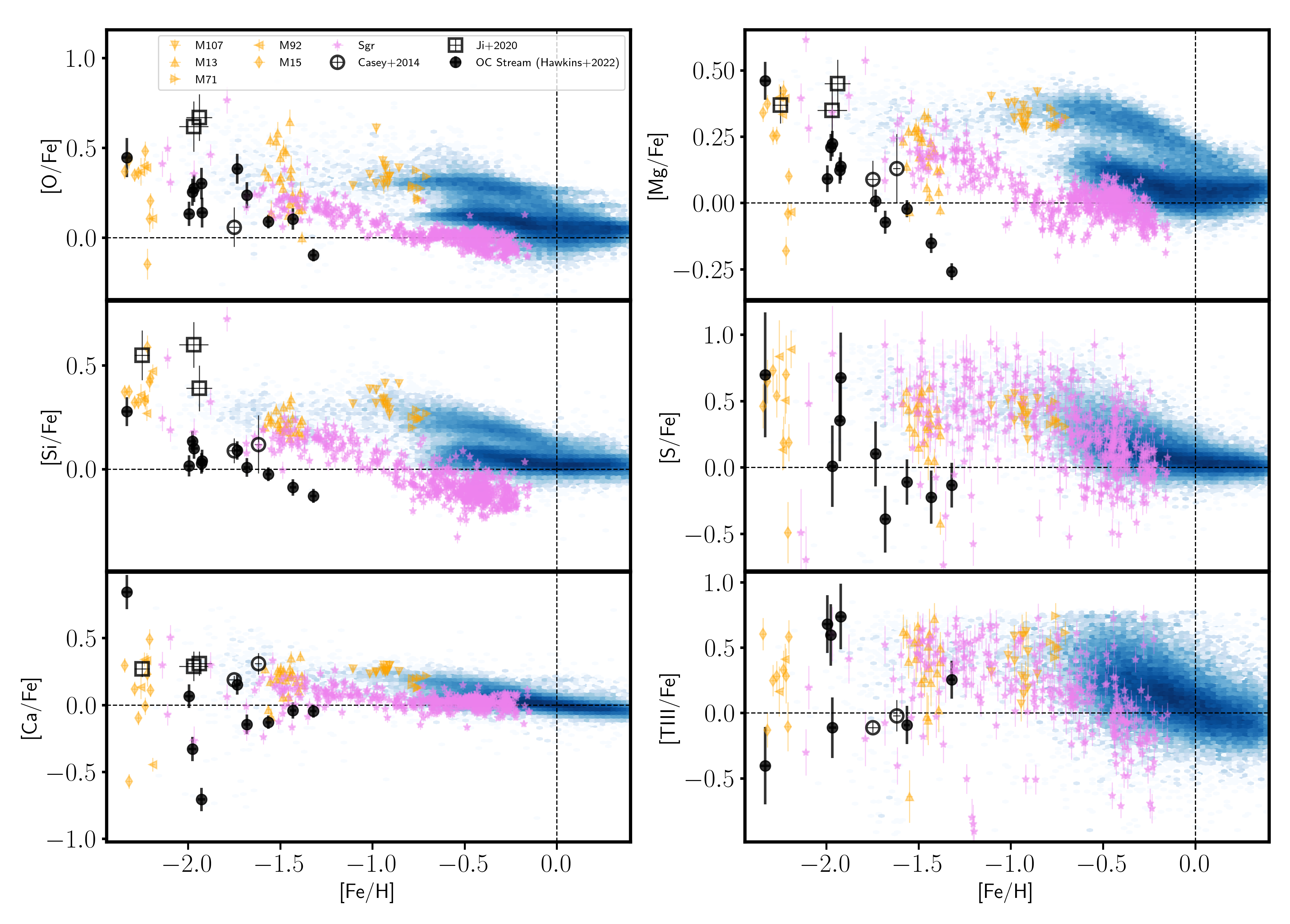}
	\caption{The $\alpha$ element abundance ratios (labeled on the vertical axis) as a function of iron abundance for OC stars observed in APOGEE (black circular markers), compared to consistently derived abundances of several stellar clusters (orange markers) including M107 (upside-down triangle), M13 (triangles), M71 (right pointed triangle), M82 (left pointed triangle, and M15 (diamonds). In addition, we also show the abundances of a Milky Way comparison sample (with $|b| < 10^{\circ}$, shown as background density in blue) and Sagittarius dwarf \protect\citep{Hasselquist2021} galaxy stars (violet colored stars) all of which are observed and analyzed homogeneously with the APOGEE survey. Finally the 3 highly probably OC stars studied in \citet[open black circles,][]{Casey2014} and \citet[open black squares,][]{Ji20} are also shown.}
	\label{fig:alpha}
\end{figure*}

The ratio $\alpha$ elements with respect to iron, i.e. [O,~Mg,~Si,~Ca,~S,~Ti/Fe], are widely used in Galactic archaeological studies because of their usefulness as a discriminator of the star-formation history and past enrichment of the gas from which stars formed \citep[e.g.,][and references therein]{Nomoto2013}. Additionally the \afe\ abundance ratio is thought to be sensitive to the ages of stars. This is because the $\alpha$ elements are largely dispersed into the interstellar medium on short time scales by core-collapse Type~II supernovae from massive stars. On the other hand, Fe and Fe-peak (Mn, Cr, Co, etc.) elements are dispersed by Type~Ia supernovae requiring a white dwarf. These occur on longer timescales \citep[e.g.][]{Gilmore1998, Matteucci2001,Maoz2012} after Type~II supernovae. Therefore, old stars will be enhanced in \afe\ but have low metallicities while younger stars (having been polluted by more Type~Ia explosions) will be depleted in the \afe\ ratio but at higher metallicity. This creates a characteristic `knee' in the \afe--\feh\ plane. The metallicity at which this `knee'  occurs is sensitive to the star formation rate (and thus mass) of the system in which the stars formed. For reference, the Milky Way has a  `knee'  in the \afe--\feh\ plane at $\feh \sim -1.0~\textrm{dex}$. This is in contrast to a much lower mass dwarf galaxy such as the Sculptor dwarf spheroidal galaxy, which have lower \afe\ at a given metallicity compared to the Milky Way and potentially has a `knee'  in the \afe--\feh\ plane at $\feh \sim -1.87~\textrm{dex}$ \citep[see Fig.~3 of][]{deBoer2014}. 

In Fig.~\ref{fig:alpha}, we show the abundance ratios of O, Mg, Si, Ca, Ti, and S with respect to Fe as a function of \feh, i.e. [O,~Mg,~Si,~Ca,~S,~Ti/Fe]. We note here that we specifically choose to plot the [Ti~II/Fe] ratio within APOGEE rather than [Ti~/Fe] due to known issues with the analysis of Ti~I lines from the H-band \citep[e.g. see discussion in][]{Hawkins2016b, Jonsson2018}. In this figure (as well as Fig.~\ref{fig:Fepeak}), the probable OC stream stars observed in the APOGEE survey (this study) are noted as black filled circles, while those observed in \cite{Casey2014} are shown as black open circles and those from \cite{Ji20} are shown as black open squares. As reference, we also show several comparison samples observed and analyzed homogeneously within the APOGEE survey. These include a Milky Way sample\footnote{As a simple Milky Way comparison sample, we select APOGEE stars with the same quality control cuts as in Section~\ref{subsec:APOGEE}. We make an additional criteria to select Milky Way stars, namely that the absolute Galactic latitude, |b|, of the stars had to be less than 10$^\circ$.} (shown as a blue-scaled background), several globular clusters (closed orange symbols), and candidate stars from the Sagittarius dwarf galaxy. The stellar clusters included are M107 (downward pointed orange triangles),  M13 (upward pointed orange triangles),  M71 (rightward pointed orange triangles), M92 (leftward pointed orange triangles),  M15 (orange diamonds).

\begin{figure}
	 \includegraphics[width=1\columnwidth]{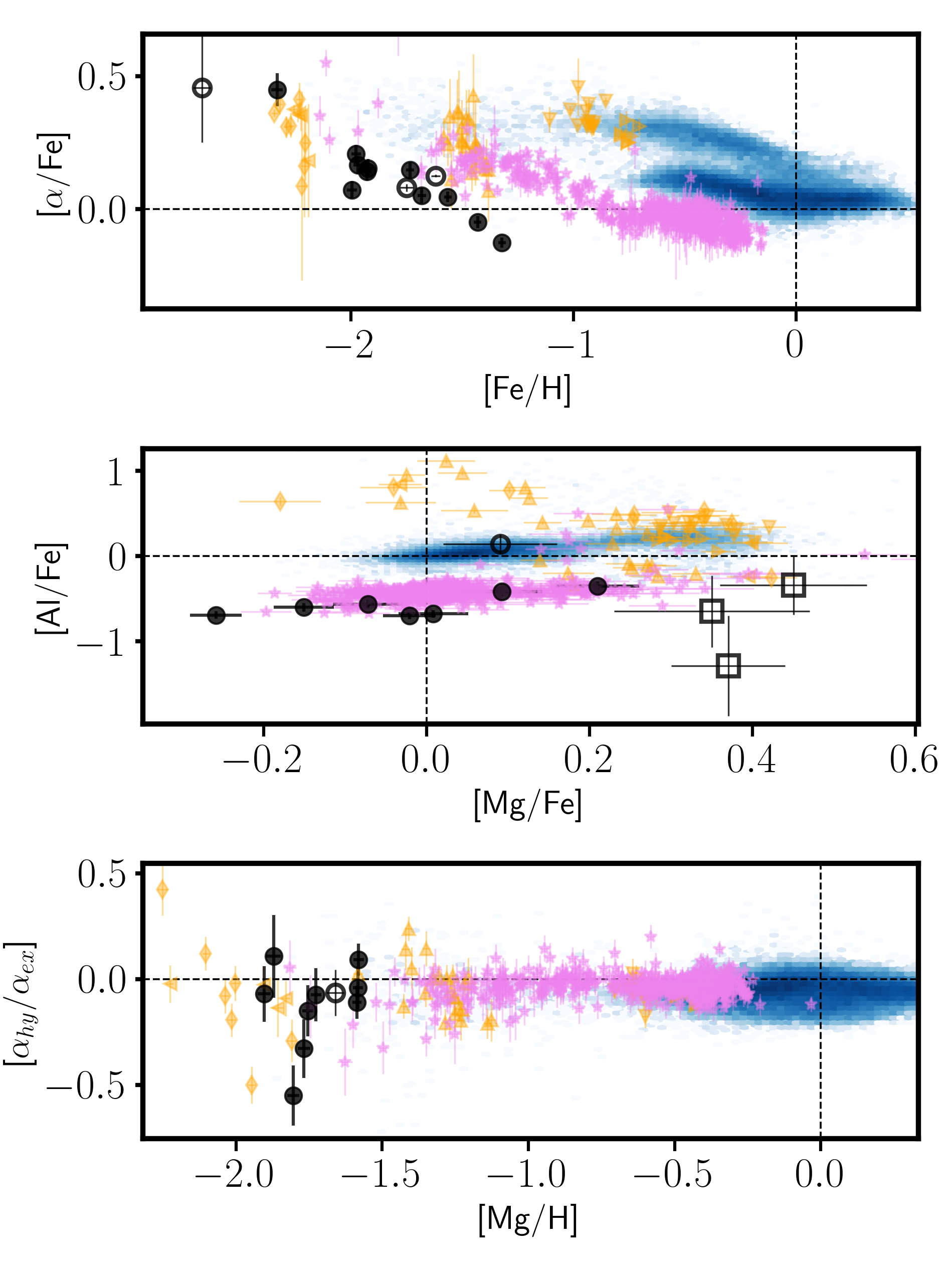}
	\caption{Top Panel : The \afe\ abundance ratio as a function of \feh. Middle panel: The [Al/Fe] as a function of [Mg/Fe] showing that the OC stream stars do not behave as one would expect from Globular cluster stars. Bottom Panel: The so called Hex abundance ratio as a function of \feh. The Hex ratio and its importance is described in the text and in \citet{Carlin2018}.   In all panels, the symbols are the same as in Fig.~\ref{fig:alpha}}
	\label{fig:alpha_HEX}
\end{figure}

It is clear from Fig.~\ref{fig:alpha} that the OC stream is {\it not} mono-metallic, as one might expect if it originates from a chemically homogeneous cluster of stars, and instead spans metallicities from $-2.10 < \feh\ < -1.50~\textrm{dex}$. This result is similar to \cite{Casey2014} and \cite{Ji20}, who both study up to 3 stars each in the OC stream. In addition, we find that generally the ratio of $\alpha$ elements, specifically O, Mg, Si, Ca, Ti, and S, with respect to iron are lower in the OC stream stars compared to homogeneously analyzed Milky Way sample. However, we note here that there is significant scatter in the [Ca/Fe] abundance of OC stream stars in the APOGEE survey, which is not seen in the other datasets \citep{Casey2014, Ji20}. This former point indicates that the stars in the OC stream were formed in an environment similar to a dwarf galaxy and not the Milky Way. In fact, taking the stars from this study and those of \cite{Casey2014} and \cite{Ji20} together, one can conclude that the \afe\ abundance ratio\footnote{There are various ways of defining (or measuring) the \afe\ abundance ratio. While APOGEE provides a globally-derived [$\alpha$/M] abundance ratio determined during the global spectral fitting procedure within the ASPCAP pipeline (i.e. the column labelled ALPHA\_M in the APOGEE catalogue), the literature studies \citep[e.g.,][]{Casey2014}, do not. Therefore, we compute, for each set of stars, a commonly-derived \afe\ by averaging the [Mg/Fe], [Si/Fe], [O/Fe] abundance ratios. The elements Mg, Si, O, were chosen because they are the best measured $\alpha$ elements in APOGEE. We note that while the OC stars in APOGEE and \cite{Casey2014}, have Mg, Si, and O measured, those in \cite{Ji20} do not have Si measurements. As a result, we do not include the three stars from \cite{Ji20} in the \afe (top panel), and HEx ratio (bottom panel) plots of Fig~\ref{fig:alpha_HEX}. } increases with decreasing metallicity with a plateau somewhere below \feh\ $<$ --2.0~dex.


The lack of a knee in the \afe--\feh\ plane (see top panel of Fig.~\ref{fig:alpha_HEX}) above $\feh \gtrapprox -2.0~\dex$ and a metallicity distribution the is peaked at \feh\ = -1.92~dex and a dispersion of 0.28~dex implies that the OC stream stars may originate in a system below the mass of Sculptor ($M \lessapprox 1.7\times10^7~\Msun$) \citep[e.g.,][]{Lianou2013}. This metallicity distribution is similar, but slightly metal-poorer, compared to earlier works \citep[e.g.][]{Li2022, Naidu2022}. Additionally, the mean metallicity of the OC stream stars imply a progenitor mass of $\sim$ 8$\times10^5~\Msun$ using the mass-metallicity relation of dwarf galaxies \citep[][]{Kirby2013}. This value is not far, but slightly lower than the predicted upper limit of $\lessapprox 9.3\times10^6~\Msun$ from \cite{Grillmair2015} and the value of $4\times10^6~\Msun$ predicted from \cite{Koposov2019}. More recently, \cite{Mendelsohn2022} estimated a total mass within 300 pc of $M = 1.1\times10^6~\Msun$, with a mass-to-light ratio of 74, for the OC stream progenitor. This mass value suggests a system much less massive than Sculptor, perhaps as low as the ultra-faint dwarf Leo IV \citep{Strigari2008}. However, we point out that while the current data implies this, we would recommend a more comprehensive study of the chemical properties of OC stream stars specifically in the metallicity range $-2.5 < \feh < -1.7~\dex$. Additional data in the regime would help pin down the exact location of any `knee' in the \afe--\feh\ diagram. We also note that while we assume here that the stream has the same metallicity of progenitor, if the OC stars were stripped from a dwarf galaxy they may be more metal-poor than the actual parent system.

As is noted in \cite{Carlin2018}, the $\alpha$ elements can actually be separated into those formed during hydrostatic carbon and neon burning (Mg, O) and those formed in the explosion event of the Type~II supernova (Si, Ca, Ti). Those authors propose a useful abundance ratio constructed by taking the average abundances of the hydrostatic (Mg, O) $\alpha$ elements relative to their explosive counterparts \citep[HEx ratio,][]{Carlin2018}. In the bottom panel of Fig.~\ref{fig:alpha_HEX}, we further study the $\alpha$ elements by inspecting this HEx ratio ([$\alpha_{\rm h/ex}$]) as a function of [Mg/H] \citep[similar to Fig.~4 of][]{Carlin2018}.  We find that the Sgr stars in APOGEE from \cite{Hasselquist2021}, which are a mix of stars in the main body and stream, show more scatter in [Mg/H] than the Sgr stream stars from \cite{Carlin2018}, but similarly show a collective trend of HEx ratios less than zero; this trend supports a scenario in which Sgr initially had very few massive stars. The HEx ratio for the OC stream stars also decrease with [Mg/H] and show a downturn at a roughly constant value of $\abunratio{Mg}{H} \sim -2$, which would correspond to a decrease in stars formed during hydrostatic burning at that (very low) level of Mg enrichment.


\subsection{Odd-Z elements (Al, K, V) } \label{subsec:oddZ}
The H-band spectra observed within the APOGEE survey allow for the measurement of several odd-Z elements including Al, K, and V. We note that there are Na lines found in the H-band spectra, however they were not well measured for each of the OC stream stars, so we choose to not report or discuss it here. While the various odd-Z elements have different production sites, they still provide useful diagnostics of the environment where the OC stream has originated from.

For example, Al, often known as a `mild $\alpha$ element', is thought to be produced (similarly to the hydrostatic $\alpha$ elements) during carbon and neon burning in massive stars, however it is also produced in the MgAl cycle in AGB stars \citep[e.g.,][]{Samland1998,  Doherty2014, Smiljanic2016}. Upon production, Al is  dispersed into the interstellar medium through both Type~II supernovae and through the winds of AGB stars. Interestingly, Al seems to be a key element for distinguishing stars originating in the second generation of stars formed in globular clusters through the Mg-Al anti-correlation \citep[e.g.,][]{Carretta2009b,Carretta2009, Meszaros2015, Pancino2017}. This particular feature, though not necessarily present in all globular clusters, manifests as a group of stars with significant enhancements of Al (often reaching values as high as $\abunratio{Al}{Fe} \sim +1.0~\dex$) and simultaneously have a depletion in [Mg/Fe]. The results for the [Mg/Fe] and [Al/Fe] abundance results can be found in Fig.~\ref{fig:alpha} and Fig.~\ref{fig:Fepeak}, respectively. However it is more useful to view the [Mg/Fe] as a function of [Al/Fe] for the purpose of looking for potential globular clusters \citep[e.g. see Fig.~3 of][]{Pancino2017}. This can be found in the middle panel of Fig.~\ref{fig:alpha_HEX}. It clearly shows that the OC stars are not enhanced in [Al/Fe] and are, in fact, significantly depleted. Additionally the [Al/Fe] is lower than expected for the various globular clusters observed in the APOGEE survey.


K is probably produced in explosive oxygen burning in Type~II supernovae. It is observationally expected to increase with decreasing metallicity, similar to $\alpha$ elements also dispersed in similar explosions \citep{Hawkins2016b}. The [K/Fe] abundance of the OC stream stars (Fig.~\ref{fig:Fepeak})  observed in APOGEE seem to have significant scatter. This large scatter is not seen in the sample observed in \cite{Casey2014} with optical spectra.

The production site for V is poorly modeled in supernovae yields. Currently, it is thought that V is produced in explosive oxygen, silicon, and neon burning  in both Type~II and Type~Ia supernovae \citep[e.g.][]{Samland1998}. Theoretically, [V/Fe] is expected to increase with decreasing metallicity. This is seen within the Milky Way sample as well as the OC stream stars. However, we find in Fig.~\ref{fig:Fepeak}  that the OC stream stars have lower [V/Fe] compared to the Milky Way or known clusters at a given metallicity .

\subsection{Iron-Peak elements (Mn, Co, Cr, Ni)} \label{subsec:Fepeak}
\begin{figure*}
	 \includegraphics[width=2\columnwidth]{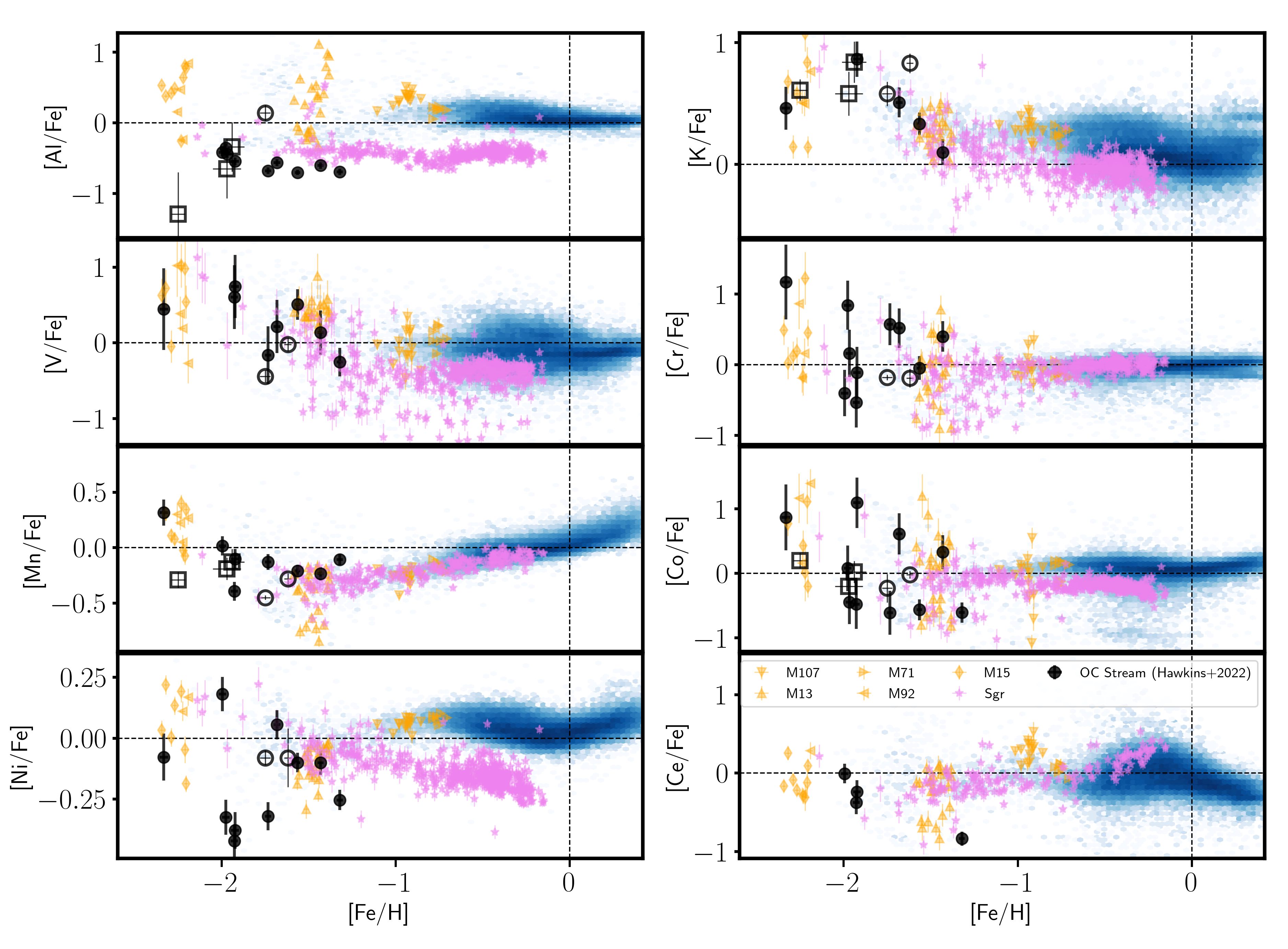}
	\caption{Left Panel : The same as Fig.~\ref{fig:alpha}, but for [Al/Fe], [V/Fe], [Mn/Fe], and [Ni/Fe] from top to bottom, respectively. Right Panel: The same as Fig.~\ref{fig:alpha}, but for [K/Fe], [Cr/Fe], [Co/Fe], and [Ce/Fe] from top to bottom, respectively. }
	\label{fig:Fepeak}
\end{figure*}

The Fe-peak elements observed within the APOGEE survey include Mn, Co, Cr, and Ni. While these elements are thought to be produced and dispersed largely through Type~Ia explosions, many of these Fe-peak elements are also produced in Type~II supernova expositions \citep[][and references therein]{Iwamoto1999, Kobayashi2006, Kobayashi2009, Nomoto2013}. The [Mn, Co, Cr, Ni/Fe] abundance ratios of the Fe-peak elements as a function of metallicity for the OC stream (and comparison sample) stars can be found in Fig.~\ref{fig:Fepeak}.

Mn is thought to be produced in significantly higher amounts in Type~Ia supernovae compared to Fe and is therefore a good tracer of such explosions. However,  unlike many of the other Fe-peak elements, it is expected (and observed) that the abundance ratio [Mn/Fe] decrease with decreasing metallicity in the Milky Way. This pattern could be due to a metallically dependence on the yields of Mn or a delay in enrichment from Type~Ia supernovae \citep[e.g.][]{Kobayashi2006, Feltzing2007}. As expected, the Milky Way (blue scale hexagonal bins in Fig.~\ref{fig:Fepeak}) the [Mn/Fe] does decrease with decreasing metallicity. We find a relatively constant sub-solar [Mn/Fe] for most OC stream stars although with significant scatter. This is in contrast to the results of \cite{Casey2014} and \cite{Ji20}, which both show a decreasing trend in [Mn/Fe] with metallicity, albeit offset from each other and with small samples (there are only two OC stream stars both optical studies which have measured Mn abundances). Further, we find that [Mn/Fe] is slightly higher in the OC stream compared to both the Milky Way and the SGR at similar metallicities.  

Cr and Ni are thought to be produced in similar ways (and levels) as Fe in both Type~Ia. Therefore it is both expected and observed that Cr and Ni track Fe in lockstep (i.e. the [Ni/Fe] and [Cr/Fe] values are fairly constant over a broad range in \feh). This is observed for both the OC stream stars and the comparison samples, albeit with moderate dispersion. We also find that [Ni/Fe] for the OC stars is slight lower compared to Milky Way and SGR stars. Recent studies of Mn and Ni abundance patterns in dwarf galaxies \citep[e.g.,][]{Sanders2021}, have shown that the sub-Chandrasekhar mass systems are a significant contribution to Type Ia supernova in metal-poor, dwarf galaxy-like environments. We find abundance patters (e.g., sub-solar [Ni/Fe] and [Mn/Fe]) that are consistent with this. 

Finally, the production of Co, while similar to Cr and Ni, is through both Type~Ia and Type~II supernovae. However unlike Cr and Ni, the amount of Co produced in the explosions  is thought to depend on both mass and metallicity \citep{Kobayashi2006} and therefore the [Co/Fe] ratio increases with decreasing metallicity for subsolar \feh. This abundance pattern is seen in the comparison samples as well the OC stream stars, which is not seen in the stars where Co could be measured in \cite{Casey2014}. We also note in all of the Fe-peak elements the OC stream stars could not really be distinguished from the comparison Milky Way sample in mean or dispersion.

\subsection{S-Process elements (Ce)} \label{subsec:sprocess}
It was noted in \cite{Hawkins2016b} that the $H$-band spectra obtain by APOGEE contain s-process information in addition to the $\alpha$, Fe-peak, and odd-Z elements that had been known before \citep[e.g.][]{Holtzman2015}. They identified Rb, Nd, and Yb but noted that each were difficult to measure at the resolution. Following this, \cite{Hasselquist2016, Hasselquist2021} identified and measured Nd in the core of the Sagittarius dwarf galaxy and \cite{Cunha2017} identified several Ce lines within the APOGEE spectra. We checked whether Ce and Nd were measured in any of the OC stream stars with the \textsc{ASPCAP} pipeline. We found in each case the Nd could not be measured in such low metallicity stars given the very weak line strength.  Ce, on the other hand, could be measured in four stars. In the bottom right panel of  Figure~\ref{fig:Fepeak}, we show the [Ce/Fe] abundance ratio as a function of \feh for OC stream stars (red circles), compared to the other Milky Way stars (blue-scaled background), globular clusters (orange symbols), and Sagittarius dwarf galaxy (magenta stars) stars. The [Ce/Fe] for values are on the lower end (all showing negative in [Ce/Fe]) compared to those found in halo field stars of similar metallicity \citep[e.g. see Fig.~7 of][]{Cunha2017}. We note however that spectra with higher signal-to-noise for many more OC stream stars will be required to draw any conclusions.

%

\section{Discussion} \label{sec:discussion}
In this section, we contextualize the chemical abundance patterns for OC stream stars and what it can tell us about what the parent population may (or may not) be. Namely, in section~\ref{subsec:GC} we evaluate the hypothesis that the OC stream was created by a fully-disrupted globular cluster. In section~\ref{subsec:dwarf}, we contrast that scenario with the idea that the OC stream was created by a disrupted dwarf galaxy.

\subsection{Ruling Out Globular Cluster Origin} \label{subsec:GC}
Globular clusters are old, metal-poor relics that contain many ($10^{4}-10^6$) stars. It has been shown that several globular clusters in the Milky Way halo are being tidally disrupted by the Milky Way \citep[e.g., the tidal tails around Palomar 5 or Palomar 13][]{Odenkirchen2003, Shipp:2020}. It is therefore possible that the OC stream could be produced by a relic globular cluster. 
\citet{Koposov2019} notes that 7 globular clusters are within 7$^{\circ}$ of the stream's great circle.

There are a handful of globular clusters that have been mentioned in the literature that could be associated with the OC stream. These include several clusters that have already been discussed in the literature as possible parents for the OC stream, namely NGC~2419 \citep[discussed in][]{Bruns2011} and Ruprecht~106 \citep[discussed in][]{Grillmair2015}, and Laevens~3 \citep{Li2022}. In the first two cases the authors conclude that while these clusters are close by to the stream \citep[Ruprecht~106 being the closest,][]{Grillmair2015, Koposov2019}, they are not likely to be associated with either, which we confirm here. 

From the perspective of chemical abundance patterns, we are in a position to evaluate the claim that the OC stream could originate from a globular cluster. First, from section~\ref{sec:results}, we show that the OC stream is not mono-metallic and has considerable scatter in many abundances. This rules out a chemically homogeneous globular cluster formation scenario. Given that Ruprecht~106 has been shown to likely be a single population, chemically-homogeneous globular cluster \citep{Villanova2013}, we rule out this cluster as a potential parent.

However, we know that not all globular clusters are mono-metallic and those which are not have second generation stars which often display anti-correlations between Mg-Al and Na-O \citep[e.g][]{Carretta2009b,Carretta2009, Meszaros2015, Pancino2017,Bastian2018}. In these clusters it is expected [Al/Fe] is enhanced while [Mg/Fe] is depleted in the second generation stars. In Fig.~\ref{fig:alpha_HEX}, we clearly show that none of the OC stream stars display such chemical signatures.

Furthermore, NGC~2419, a cluster said to be a potential parent to the OC stream \citep{Bruns2011}, has been shown to uniquely have an anti-correlation between [K/Fe] and [Mg/Fe] \citep{Cohen2011, Mucciarelli2012}. These authors found that NGC~2419 has stars which have simultaneously high K (with $\abunratio{K}{Fe} > 1$) and low Mg (with $\abunratio{Mg}{Fe} < 0$). Similar to \cite{Casey2014}, we find that the [K/Fe] is too low to be consistent with this globular cluster. We therefore rule out this cluster as a potential parent.

Finally, Laevens~3 was proposed by \cite{Li2022} as a potential parent progenitor of the OC stream. Only one detailed chemical study of the Laevens~3 cluster exists \citep{Longeard2019}, which explore the metallicity and \afe\ of a total of 3 member stars. The metallicity they derive for the cluster is \feh\ = --1.8 $\pm$0.10~dex and a metallicity dispersion of $\sigma$\feh\ $<$ 0.50~dex. Additionally, they find a \afe\ = 0.00 $\pm$ 0.20~dex. The APOGEE OC stream stars have a median metallicity of -1.92 with a dispersion of $\sim$0.28~dex. While the metallicity is consistent with those of Laevens~3 the mean \afe\ of Laevens~3 is 0.00 dex, while the OC stream stars have a significantly higher mean \afe\ of 0.15~dex. Given this we note that Laevens~3 is not a likely parent of the OC stream. However we note that a more detailed chemical characterization of Laevens~3 and the OC stream would be helpful to further illustrate this.

Given that none of the globular clusters that are close to the OC stream match its phase-space tracks \citep{Koposov2019} as well as the chemical abundance information presented here and in \cite{Casey2014}, the lack of an Mg-Al or K-Mg anti-correlation, and the spread in \feh, we determine that it is unlikely that the parent of  OC stream is a globular cluster. 

\subsection{Dwarf Galaxy Origins for the Orphan Stream} \label{subsec:dwarf}

In the last section, consistent with results on a smaller sample of stars from  \cite{Sesar2013} and \cite{Casey2014}, we conclude that the OC stream could not have originated from a disrupting globular cluster. Therefore, in this section, we evaluate the possibility that the OC stream has originated from a dwarf galaxy system (including ultra faint and classical dwarfs). In \cite{Koposov2019}, it is noted that there are 3 ultra faint dwarf galaxies (Segue~1, Ursa~Major~II, and Grus~II) and 1 classical dwarf spheroidal galaxy (Leo~I) in the vicinity of the OC stream. Of these Leo~I is automatically ruled out because it is much farther away ($\sim$250~kpc) compared to the stream and thus they are not likely physically associated \citep{Koposov2019}. We do not discuss Leo~I further for this reason.

In the recent works \citep[e.g.,][]{Casey2014}, it was discussed that the OC stream progenitor could be the Segue~1 dwarf spheroidal. With the chemical characterization in this work, we are in a position to assess this as a possibility. The chemical patterns of the Segue~1 dwarf spheroidal were studied in \cite{Frebel2014}. In that work, they indicated that Segue~1 is one of the only dwarf galaxy systems where the ratio of [Mg, Si, Ca/Fe] are actually enhanced and show no evidence for a `knee'. This is inconsistent with the abundance patterns seen in section~\ref{subsec:alpha}, where we find that the [Mg, Si, Ca/Fe] abundance ratios are not only depleted relative to the rest of the Milky Way halo but they also increase with decreasing metallicity indicating there is likely a `knee' below $\feh < -2~\dex$. Assuming that the chemical pattern observed in the OC stream should be drawn from the chemical make-up of its progenitor, our results rule out Segue~1 a possible progenitor to the OC stream.

Ursa~Major~II has also been discussed as a potential parent of the OC stream \citep{Fellhauer2007}. \cite{Kirby2008} was among the first to derive chemical abundance information, though only \feh. They found in that work that Ursa~Major~II has a mean metallicity of $\feh\ \sim -2.44~\dex$ with a spread of 0.57~dex. The first detailed chemical abundance study of  Ursa~Major~II was done by \cite{Frebel2010}; However, in that work they only obtained chemical abundances for three stars and each of those stars happened to have metallicities $\feh < -2.3~\dex$ (significantly lower than the \feh\ of the OC stream stars discussed in this work).

Another dwarf galaxy has been recently presented \citep[e.g.][]{Koposov2019, Erkal2019} as a potential parent of the OC stream, namely the Grus~II system. The Grus~II system is an ultra faint dwarf galaxy at distance of $\sim$49~kpc from the Galactic center and an estimated physical size of $\sim$93~pc \citep{Drlica-Wagner2015}. To date, the chemical pattern of Grus~II is still poorly constrained. \cite{Hansen2020} completed the first and only detailed chemical study of three of the brightest members of Grus~II, which were in the metallicity range $-2.95 < \feh < -2.49~\dex$. If this metallicity range is representative of Grus~II, it would indicate that the either Grus~II has a very large metallicity spread or that the OC stream is likely too metal-rich to have originated within Grus~II.  Further chemical characterization of Grus~II as a more metal-rich (i.e., $\feh > -2.10~\dex$) dwarf galaxy would be a necessary step to determine if it is connected to the OC stream. With limited chemical abundance information from Grus~II, Ursa~Major~II, and the OC stream, we recommend detailed chemical abundance studies in overlapping metallicity regimes for each in order to distinguish which could be the parent.   We here that all of the dwarf galaxies listed here are likely offset from the OC stream in at least one component of its track \citep[e.g.,][]{Koposov2019} or the chemical pattern does not match, which illustrates that the hunt for a progenitor is still ongoing. The chemical patterns presented in this work will help identify a progenitor.

\section{Summary}  \label{sec:summary}
The Orphan--Chenab (OC) stream is an relatively-old, fairly narrow stream of stars found around the Milky Way \citep[e.g.][]{Grillmair2006, Belokurov2007} with no known progenitor system. Recently, the kinematics of the OC stream has been studied and modeled using data from the second data release of the all-sky \gaia\ survey \citep{Koposov2019, Erkal2019, Fardal2019}. These studies have furthered our knowledge of the spatial and dynamical properties of this interesting. An interesting but still rather unexplored avenue to help find the parent of the OC stream lies in matching its detailed chemical abundance pattern to possible progenitor systems.

Only two studies currently exists on the detail abundances of at most 6 probable OC stream members \citep{Casey2014,Ji20}. In this work, we use the updated kinematic and spatial properties of the OC stream to find 13 probable OC stream stars within the APOGEE survey. The APOGEE survey contains moderate resolution, H-band spectra for more than a few times 10$^5$ stars. These spectra enable the atmospheric (\teff, \logg) and chemical characterization of the stars in various elemental families including the $\alpha$ (O, Mg, Ca, Si, Ti, S), odd-Z (Al, K, V), and Fe-peak (Fe, Ni, Mn, Co, Cr) families. Of the 14 probable OC stream stars,  5 have measured stellar parameters and chemical abundances in the APOGEE survey. This represents the largest study of the chemical abundance pattern of the OC stream and the first with a large spectroscopic survey in the infrared.

Our results can be summarized as follows: \begin{enumerate}
\item The metallicity of the OC stream ranges at least from $-2.10 < \feh < -1.55~\dex$ indicating that is not mono-metallic, consistent with other findings \citep{Casey2014}. We also find that in many of the elements studied, that the OC stream is not a mono-abundance population.

\item The $\alpha$ elements are largely depleted compared to the Milky Way reference sample at similar metallicities (Fig.~\ref{fig:alpha} and top panel of Fig.~\ref{fig:alpha_HEX}). This result is particularly important because it indicates that the stars that make up the OC stream have not likely originated in the Milky Way. Instead, they have come from an environment with lower star formation rates (e.g., dwarf spheroidal or ultra faint dwarf galaxies.

\item The distribution of OC stars in the \afe--\feh\ abundance plane seems to indicate that the OC stream has no reasonable `knee' at \feh\ larger than $-2.0~\dex$, implying an upper limit on the mass of its progenitor of $M \lessapprox 1.4\times10^7~\Msun$). We emphasize, however, that many more stars with $\feh\ < -2~\dex$ are needed to pin down if and at what metallicity a `knee' can be found. The metallicity distribution combined with the \afe\ abundances suggest the OC progenitor should be between $8\times10^5 \lessapprox$ M $ \lessapprox 1\times10^7  ~\Msun$

\item Studying the abundance patterns of Mg and Al (in Fig.~\ref{fig:alpha_HEX}) indicates that stars making up the OC stream stars {\it are not consistent with originating in a known globular cluster} consistent with other findings \citep[e.g.][]{Casey2014, Ji20}.

\item  We find that the dispersion in many Fe-peak elements (Mn, Cr, Ni) is significantly larger than what is found with 3 OC stream candidate stars from \cite{Casey2014} and those of \cite{Ji20}. This discrepancy underscores the need for significantly larger samples of stars within the OC stream in order to further constrain the nature of its chemical pattern elements. Interestingly, we also find that the OC stream is offset in some Fe-peak elements (e.g., Ni, Mn) compared to the Milky Way. 
\end{enumerate}

We are currently working on two paths forward based on these results. Firstly, we need to further evaluate the claim the the Grus~II dwarf spheroidal could be a potential parent for the OC stream. In order to properly test this, we are working to obtain high-resolution spectra of tens of stars within Grus~II to determine if its metallicity and chemical abundance patterns are consistent with the OC stream. An initial chemical study of 3 stars in Grus~II was carried out by \cite{Hansen2020}. These authors found that Grus~II is very metal-poor (i.e., $-2.49 < \feh < -2.94~\dex$) and has a chemical pattern (elevated [Mg/Ca] ratios) that suggest it likely had a top-heavy initial mass function. It is clear that a larger study of the system focusing on candidates at higher metallicity (one that overlaps with the OC) stream is required. Further, kinematic and dynamical studies of all potential progenitors that include the impact of the Large Magellanic Cloud \citep[e.g.,][, Koposov et al., in preparation]{Lilleengren2022}, should be done in order to confirm any potential progenitor of the OC stream. While in this work, we further advance our understanding of the chemical nature of the OC stream by increasing the sample size by a factor of 2, we still have only relatively small numbers. Therefore, we also recommend exploring the chemical pattern of 24 OC stream candidate found in the LAMOST survey \citep[e.g.][]{Li2017} using advanced spectral analysis techniques such as \thepayne\ \citep{Ting2017}.
%
%
%
%
%
%

\section*{Acknowledgements}
{\small
KH acknowledges support from the National Science Foundation grant AST-1907417 and AST-2108736 and from the Wootton Center for Astrophysical Plasma Properties funded under the United States Department of Energy collaborative agreement DE-NA0003843. This work was performed in part at Aspen Center for Physics, which is supported by National Science Foundation grant PHY-1607611. This work was performed in part at the Simons Foundation Flatiron Institute's Center for Computational Astrophysics during KH's tenure as an IDEA Fellow.

AAS and AZS acknowledge support from the SDSS FAST Initiative (\url{https://www.sdss.org/education/faculty-and-student-team-fast-initiative/}), in particular funding for the CUNY FAST group.

For the purpose of open access, the author has applied a Creative Commons Attribution (CC BY) license to any Author Accepted Manuscript version arising from this submission.

This work has made use of data from the European Space Agency (ESA)
mission {\it Gaia} (\url{https://www.cosmos.esa.int/gaia}), processed by
the {\it Gaia} Data Processing and Analysis Consortium (DPAC,
\url{https://www.cosmos.esa.int/web/gaia/dpac/consortium}). Funding
for the DPAC has been provided by national institutions, in particular
the institutions participating in the {\it Gaia} Multilateral Agreement.

Funding for the Sloan Digital Sky 
Survey IV has been provided by the 
Alfred P. Sloan Foundation, the U.S. 
Department of Energy Office of 
Science, and the Participating 
Institutions. 

SDSS-IV acknowledges support and 
resources from the Center for High 
Performance Computing  at the 
University of Utah. The SDSS 
website is www.sdss.org.

SDSS-IV is managed by the 
Astrophysical Research Consortium 
for the Participating Institutions 
of the SDSS Collaboration including 
the Brazilian Participation Group, 
the Carnegie Institution for Science, 
Carnegie Mellon University, Center for 
Astrophysics | Harvard \& 
Smithsonian, the Chilean Participation 
Group, the French Participation Group, 
Instituto de Astrof\'isica de 
Canarias, The Johns Hopkins 
University, Kavli Institute for the 
Physics and Mathematics of the 
Universe (IPMU) / University of 
Tokyo, the Korean Participation Group, 
Lawrence Berkeley National Laboratory, 
Leibniz Institut f\"ur Astrophysik 
Potsdam (AIP),  Max-Planck-Institut 
f\"ur Astronomie (MPIA Heidelberg), 
Max-Planck-Institut f\"ur 
Astrophysik (MPA Garching), 
Max-Planck-Institut f\"ur 
Extraterrestrische Physik (MPE), 
National Astronomical Observatories of 
China, New Mexico State University, 
New York University, University of 
Notre Dame, Observat\'ario 
Nacional / MCTI, The Ohio State 
University, Pennsylvania State 
University, Shanghai 
Astronomical Observatory, United 
Kingdom Participation Group, 
Universidad Nacional Aut\'onoma 
de M\'exico, University of Arizona, 
University of Colorado Boulder, 
University of Oxford, University of 
Portsmouth, University of Utah, 
University of Virginia, University 
of Washington, University of 
Wisconsin, Vanderbilt University, 
and Yale University.

}

\bibliographystyle{aasjournal}
\bibliography{bibliography}
\label{lastpage}
\end{document}